\documentclass[floatfix,10pt,superscriptaddress,twocolumn,showpacs,showkeys,aps,prb,groupedaddress,final]{revtex4-2}
\usepackage{amsmath}
\usepackage{amsfonts}
\usepackage{amssymb}
\usepackage{xcolor}
\usepackage{graphicx}%
\def\6{\partial}

\usepackage{ifpdf,braket}
\ifpdf
\usepackage[pdftex,
        colorlinks=true,
        citecolor=blue,
        linkcolor=blue,
        urlcolor=blue,
        pdftitle={The intermediate type-I superconductors in the mesoscopic scale},
        baseurl={http://www.df.ufrpe.br},
        pdftoolbar=true,
        pdfmenubar=false,
        pdfpagemode=UseOutlines,
        pdfstartview=FitH,
        bookmarksopen=false
        ]{hyperref}
\fi
\begin{document}

%\preprint{APS/123-QED}

\title{The intermediate type-I superconductors in the mesoscopic scale}% Force line breaks with \\
\author{Leonardo R. Cadorim}
\affiliation{ Departamento de F\'{\i}sica, Faculdade de Ci\^encias, Universidade Estadual Paulista (UNESP), Caixa Postal 473,
17033-360, Bauru-SP, Brazil }
%\email{edson.sardella@unesp.br }
\author{Antonio R. de C. Romaguera}%
\affiliation{Departamento de F\'{\i}sica, Universidade Federal Rural de Pernambuco, 52171-900 Recife, Pernambuco, Brazil}
 \author{Isa\'{\i}as G. de Oliveira}%
\affiliation{Departamento de F\'{\i}sica, Universidade Federal Rural do Rio de Janeiro, CEP 23890-000, Seropédica, RJ, Brazil}
\author{Rodolpho R. Gomes}
\affiliation{Instituto de Qu\'{\i}mica, Universidade Federal do Rio de Janeiro, 21941-972 Rio de Janeiro, Brazil}
\author{Mauro M. Doria}
%\email[Corresponding author: ]{mmd@if.ufrj.br}
\affiliation{Instituto de F\'{\i}sica, Universidade Federal do Rio de Janeiro, 21941-972 Rio de Janeiro, Brazil}%
\affiliation{Instituto de F\'{\i}sica ``Gleb Wataghin'', Universidade Estadual de Campinas, 13083-970, Campinas, S\~ao Paulo, Brazil}

\author{Edson Sardella}
\email[Corresponding author: ]{edson.sardella@unesp.br}
\affiliation{ Departamento de F\'{\i}sica, Faculdade de Ci\^encias, Universidade Estadual Paulista (UNESP), Caixa Postal 473,
17033-360, Bauru-SP, Brazil }

\date{\today}% It is always \today, today,

\begin{abstract}
M.\ Tinkham and P.\ G.\ de Gennes,  described in their books\cite{tinkham04,degennes66} the existence of an \textit{intermediate} type-I superconductor as a consequence of  an external surface that affects the well known classiﬁcation of superconductors into type-I and II.
Here we consider the mesoscopic superconductor where the ratio volume to area is small and the effects of the external surface are enhanced.
By means of the standard Ginzburg-Landau theory the Tinkham-de Gennes scenario is extended to the mesoscopic type-I superconductor. We find new features of the transition at the passage from the \textit{genuine} to the \textit{intermediate} type-I. The latter has  two distinct transitions, namely, from a paramagnetic to diamagnetic response in descending field and a \textit{quasi} type-II behavior as the critical coupling $1/\sqrt{2}$ is approached in ascending field.
The intermediate type-I phase proposed here, and its corresponding transitions, reflect intrinsic features of the superconductor and not its geometrical properties.
\end{abstract}

\maketitle

%\tableofcontents
\par
\section{Introduction}
The classification of superconductors on the basis of their magnetic properties has been a hard earned  knowledge.
Nearly twenty five years after  the  discovery of superconductivity by H. Onnes, the experimental measurements of  L. Shubnikov  showed that the magnetic properties of alloys were very different from those of the  pure metals~\cite{shepelev10}.
The explanation had to wait for more twenty years until the development of the theoretical work  of A. Abrikosov~\cite{abrikosov56}, based on a new and unknown at the time phenomenological theory, the Ginzburg-Landau (GL) theory.
Nowadays the GL theory enjoys enormous recognition for its applications in various fields, ranging from phase transitions to particle theory since the  Higgs model may be regarded as a  relativistic generalization of the GL theory~\cite{nielsen73}.
The classification of superconductors is straightforwardly obtained from the ratio between two fundamental measurable lengths, namely, the London penetration length ($\lambda$) and the coherence length ($\xi$).
A. Abrikosov found that the single coupling of the GL theory,  $\kappa \equiv \lambda/\xi$, splits the superconductors into two classes, namely, type-I and II, and the  critical value separating them is $\kappa=1/\sqrt{2}$.
Although his simplified geometry is beyond reality since there are no boundaries, this choice is useful in the sense that excludes any geometrical factor from entering the classification scheme.
This critical coupling was also found by E. Bogomolny~\cite{bogomolny76,lukyanchuk01} in the context of string theory thus rendering the transitions in $\kappa$ obtained here of possible interest to other areas of physics besides superconductivity.
The  magnetic difference between type-I and II stems from  the existence of a vortex state in the type-II that disappears when the normal state sets in at the upper critical field $H_{c2}$. The type-I simply does not sustain a vortex state and goes to
the normal state at the thermodynamic field $H_c$, where the  normal and the superconducting  Gibbs' free energies become equal.
Superconductivity was discovered in the pure elements, known to be type-I with the exception of  Nb, V, and Tc which are type-II. Distinctively, from Shubnikov's days to now, superconductivity has been discovered in alloys and other composite materials~\cite{webb15}, which are mostly type-II. The many new families of high-$T_c$ superconductors~\cite{wesche17} fall in the latter case, such as the cuprates, fullerenes, MgB$_2$, pinictides and many others. Nevertheless unexpected type-I superconductivity has been found in some alloys, such as  TaSi$_2$~\cite{gottlieb92}, the heavily boron-doped silicon carbide~\cite{kriener08}, YbSb$_2$~\cite{zhao12}, and more recently in the ternary intermetallics YNiSi$_3$ and LuNiSi$_3$~\cite{avila19}, rendering them of great interest, and  worth of study.
The coupling $\kappa$ varies greatly among  superconductors ranging  from very high,  for the high-T$_c$ materials YBaCuO$_{7-\delta}$ ($\kappa =95$), to very low values, for the pure metals~\cite{wesche17} (Al (0.03), In (0.11), Cd (0.14), Sn (0.23) and  Ta  (0.38)) and also for some of the alloys~\cite{zhao12,avila19}, such as (YbSb$_2$ (0.05), YNiSi$_3$ (0.1)).
Since $\xi$ decreases with disorder, doping the material by impurities can produce adjustable $\kappa$ compounds that allow for the study of transitions in this coupling.

According to M. Tinkham~\cite{tinkham04} and P. G. de Gennes~\cite{degennes66} surface states split type-I superconductors into
\textit{genuine} and \textit{intermediate} classes that comprise the coupling ranges $\kappa < \kappa_c$  and $\kappa_c < \kappa < 1/\sqrt{2}$, respectively, where $\kappa_c = 0.417$.
Although this transition is driven by surface effects it solely reflects an intrinsic property of the superconductor, namely, $\kappa$, as discussed below.
There has been an intense  search for this \textit{genuine-intermediate} transition both theoretically~\cite{feder67,park67} and experimentally~\cite{buchanan65,mcevoy67,paco72,blot78} in the seventies, but after this period it became an elusive topic. The existence of the transition was even questioned~\cite{zharkov05} and its study no longer pursued. Now that type-I alloys were found,  the classification of superconductors acquired a renewed interest. In this paper  we pursue this study  in the mesoscopic scale, which offers a unique framework  since surface effects are enhanced there.
In the bulk (no surface)  the relation $H_{c2}=\sqrt{2}\kappa H_{c}$ elucidates the difference between types I and II since for $\kappa < 1/\sqrt{2}$ ($\kappa > 1/\sqrt{2}$)
$H_{c2} < H_c$ ($H_{c2} > H_c$), thus a type-I (type-II) superconductor.
Interestingly M.\ Tinkham has stressed in his book~\cite{tinkham04} that both fields $H_{c2}$ and  $H_c$ are directly measurable in type-I superconductors.
The normal state is retained  below the thermodynamic field $H_c$ until the lower field $H_{c2}$ is reached~\cite{mcevoy67}, where the order parameter (magnetization) abruptly becomes non-zero.
From the other side evolving from the superconducting state this lasts beyond $H_{c2}$ until $H_{c}$ is reached and the normal state is recovered.
However the presence of a surface  modifies the above bulk analysis since superconductivity is  extended beyond $H_{c2}$ to exist  with thickness $\xi$   around the external boundary.
Saint-James and de Gennes~\cite{saintjames63} found the critical field $H_{c3} > H_{c2}$ whose value in case of a flat interface  is  $H_{c3} = 1.695H_{c2}$~\cite{saintjames63,saintjames65,kwasnitza66,hauser74,askerzade03,changjan17,xie17}.
$H_{c3}$ is also present in type-I superconductors~\cite{christiansen68} and signals several processes, e.g.,  the expulsion of magnetic flux~\cite{jung02,valko07}. In fact it is $H_{c3}$ which gives rise to the \textit{genuine} and \textit{intermediate} type-I superconductors,  associated to $H_{c2} < H_{c3} <  H_c$  and  $H_{c2} < H_c < H_{c3}$, respectively, such that $\kappa_c$ is obtained from $H_{c3}=H_{c}$.

In this paper, we  report properties of the \textit{genuine-intermediate}  transition such as its critical $\kappa$ and also other transitions in $\kappa$ in the \textit{intermediate} phase, as seen by isothermal magnetization $M(H)$ curves.
These several transitions can be experimentally investigated by the ballistic Hall magnetometry technique~\cite{geim97} applied to submicron size superconductors.
Type-I mesoscopic superconductors have been investigated both theoretically~\cite{golib09} and experimentally~\cite{simon12}, however the \textit{genuine-intermediate} transition in $\kappa$ is firstly considered here and found to acquire new properties. As one goes from the  macroscopic to the mesoscopic scale, $\kappa_c$ gives rise to $\kappa_{c1}$ and also to the transitions $\kappa_{c2}$ and $\kappa_{c3}$ inside the \textit{intermediate} phase.
In descending field starting from the normal state the magnetization can display paramagnetic regions
but it becomes diamagnetic at any applied field provided that $\kappa < \kappa_{c2}$. Hereafter we call this as the {\it dia-para} transition.
In ascending field the magnetization ($-M$) has a nearly linear growth (Meissner state) up to a maximum and next undergoes an abrupt fall and a residual magnetization regime is reached that only exists if $\kappa> \kappa_{c3}$. The Meissner state is followed by the disappearance of the magnetization for $\kappa< \kappa_{c3}$.
This residual magnetization signals the {\it quasi} type-II class and is caused by giant  vortices
(for a discussion about giant vortices in mesoscopic
superconductors see for instance
Refs.~\onlinecite{romaguera2007,baelus2006}).
We remark the presence of several notable fields in our study.
In the up branch there are $H^{\prime}_{c}$ (the peak of $-M$) and $H^{\prime\prime}_{c}$ (the vanishing of  the magnetization).
They are very near to each other for $\kappa< \kappa_{c3}$ but not for $\kappa> \kappa_{c3}$.
Both critical fields fall above $H_{c}$ (see supplementary material), and so are inside a region of metastability since the superconducting state there has higher Gibbs free energy than the normal state.
In the descending branch we define  $H'_{c3}$ where the magnetization becomes non-zero and the superconducting state sets in.
As shown here these fields, as well as the $M(H)$ curves, are strongly dependent on $\kappa$ in the type-I domain.
We bring numerical evidence that the {\it genuine-intermediate} mesoscopic transition takes place at a coupling lower than the macroscopic one, $\kappa_{c1} < \kappa_c$.
Interestingly the {\it dia-para} transition occurs at $\kappa_{c2}\approx \kappa_c$, thus near to the macroscopic transition a so far  fortuitous coincidence.
The transition to the {\it quasi} type-II class takes place for $\kappa_{c3}<1/\sqrt{2}$.

Our numerical analysis was carried on a very long needle with square cross section of size $L^2$ in the presence of  an applied field parallel to its major axis. The needle is sufficiently long such that the top and the bottom surfaces can be ignored and just a transverse two-dimensional cross section needs to be considered.
The square cross section is the most suited to our numerical procedure which is  done on a square grid. The boundary conditions are smoothly implemented in this geometry, namely, of no current exiting the superconductor and that at the surface the local field meets the external applied field. Nevertheless our major findings hold  independently of the selected cross section geometry, though the value of the critical fields and of the delimiting $\kappa_{ci}$ may be affected by it.
We look at several cross section sizes, namely $L=\rho \lambda$, $\rho=8,\,12,\,16,\,24, \;\mbox{and} \, 32$.

It is well-known that the GL theory is the leading term of an order parameter expansion derived from the microscopic BCS
theory~\cite{degennes66,arkady11,vagov16}. In case the next to leading order corrections are included~\cite{vagov16}  an intermediate phase emerges in the diagram $\kappa$ versus $T$ in between the type-I and II domains. However the analysis of A. Vagov et al.~\cite{arkady11,vagov16} does not take into account surface effects whereas here the intermediate phase is solely due to these surface effects~\cite{isaias17}, and for this reason the intermediate phase is found already in the standard GL theory level.
The choice of an infinitely long system, instead of being a limiting factor, really expands the importance of the present results, once it allows to see intrinsic effects. Geometrical factors~\cite{camacho2016,camacho2018,cadorim2019,camacho2020} hinder the observation of the intrinsic transitions observed here. It is well known that a sufficiently thin type I superconductor turns into a type II one by a change of its thickness. The geometry of the cross section affects the $\kappa$ values where transitions take place but not their existence, that only reflects the ordering among the critical fields.
\\

\section{Theoretical Formalism}
The basis of our dimensionless treatment of the GL theory is $\lambda$ and  $H_{c2}=\Phi_0/2\pi\xi^2$, that renders the free energy,
\begin{eqnarray}
    G & = & \int \, \left [ \left
    |\left ( -\frac{i}{\kappa}\mbox{\boldmath $\nabla$}-{\bf A}
    \right ) \psi\right|^2-|\psi|^2+\frac{1}{2}|\psi|^4 \right ]\,d^3r \nonumber \\
    & &  +\int \, ({\bf h}-{\bf H})^2\,d^3r\,,\label{eq:eq1}
\end{eqnarray}
in reduced units. Lengths are in units of  $\lambda$; the order parameter $\psi$  is in units of $\psi_\infty=\sqrt{\alpha/\beta}$,  where $\alpha$ and $\beta$ are the two phenomenological constants of the GL theory; magnetic fields are in units of $\sqrt{2}H_c$; and the vector potential $\textbf{A}$ is in units of $\sqrt{2}\lambda H_c$. The GL equations become,
\begin{eqnarray}\label{eq:gl}
     & -\left ( -\frac{i}{\kappa}\mbox{\boldmath $\nabla$}-{\bf A}
    \right )^2\psi+\psi\left (1-|\psi|^2\right ) = 0\;, \\
    & \mbox{\boldmath $\nabla$}\times{\bf h} = {\bf J_s}\;,
\end{eqnarray}
\noindent
where ${\bf J_s}={\rm I\!R}\left [\bar{\psi}\left ( -\frac{i}{\kappa}\mbox{\boldmath $\nabla$}-{\bf A}
    \right )\psi\right ]$ is the superconducting current density.

The GL equations were solved numerically within a suitable
relaxation method and upon using the link-variable method
as presented in Ref.~\onlinecite{gropp1996}. For this, we used a
mesh-grid size with  $\Delta x=\Delta y=0.2\lambda$.
The applied magnetic field is adiabatically
increased in steps of $\Delta H = 10^{-3} \sqrt{2}H_c$ for both up and down branches of the field. In each simulation, $\kappa$ was held fixed and the stationary state at $H\mp\Delta H$ was used as the initial state for $H$ for up and down cycles, respectively.
\\
The emergence of superconductivity in a long mesoscopic cylindrical ($R \sim \lambda$) in presence of an applied
external field has been studied by G.F. Zharkov et al.~\cite{zharkov00,zharkov01,zharkov02,zharkov03}.
Their approach is  limited to the search of solutions of the Ginzburg-Landau differential equations with  radial symmetry which limits the search to central vortex states. In our numerical search through the link variable method we find the presence of point vortices forming various geometrical patterns inside the superconductor that fall beyond their description. 
Hence the observation of the present  $\kappa_{ci}$ transitions are beyond the scope of their framework since they are limited to a sub set of the possible vortex states.

\section{Results and Discussion}
The \textit{intermediate} and the \textit{genuine} type-I classes are distinguishable by their magnetic properties. In decreasing field the \textit{genuine} class features a direct and abrupt change from the normal state to the Meissner state, \textit{i.e.}, vortices are never trapped inside the superconductor. In the same situation the \textit{intermediate} class displays vortices, which are trapped inside either as single or giant ones and then are gradually or suddenly expelled. The two classes are associated to specific $\kappa$ ranges, and to determine them we have performed a series of numerical simulations varying $\kappa$
in steps of $\Delta\kappa=0.01$. Within this precision we were  able to numerically obtain $\kappa_{c1}$, $\kappa_{c2}$ and $\kappa_{c3}$ for all the $L$'s under investigation.
In what follows, we report properties of the  \textit{genuine-intermediate}, \textit{dia-para} and \textit{quasi} type-II transitions, the latter two being inside the \textit{Intermediate} type-I class.
%----------------------------------------------------------------
\begin{figure}
    \centering
    \includegraphics[width=0.48\textwidth]{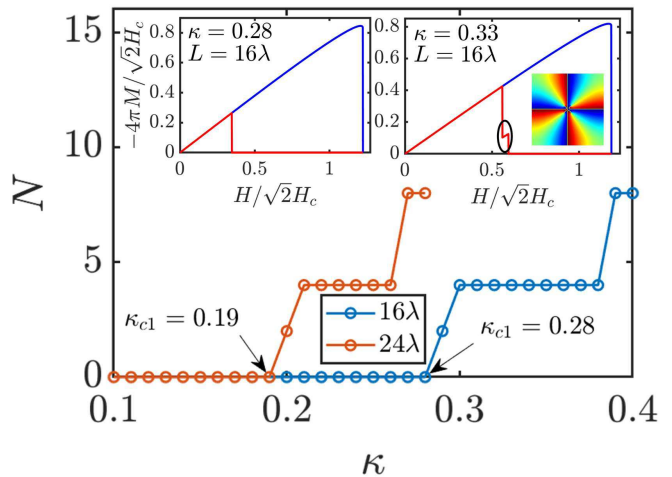}
    \caption{(Color online) The vorticity $N$  at $H^{\prime}_{c3}$, the field where the superconducting state sets in decreasing field, is plotted as a function of $\kappa$. The transition $\kappa_{c1}$ corresponds to $N=0$ marking the onset of  the \textit{Genuine} type-I class and the end of the
    \textit{Intermediate} type-I.
    The insets show typical magnetization curves for the up (blue) and down (red) branches, above and below this transition. The black ellipse highlights the spike state, which is the last possible vortex state in the \textit{Intermediate} type-I class. In the spike state a vortex nucleates and is immediately spelled from the superconductor.}
    \label{fig:fig1}
\end{figure}

Fig.~\ref{fig:fig1} features  the \textit{genuine-intermediate} transition through  the number of vortices, $N$, trapped  at $H^{\prime}_{c3}$.
The superconductor response is markedly distinct according to  $\kappa$, and this is exemplified here  for $L=16\lambda$ and $L=24\lambda$.
Above $\kappa_{c1}$, which is equal to $0.28$ for $L=16\lambda$, and $0.19$ for $L=24\lambda$, $N$ varies according to  $\kappa$,  thus corresponding to the \textit{intermediate} type-I class.
However, below $\kappa_{c1}$, $N$ drops to zero showing that no vortex enters the needle at $H^{\prime}_{c3}$, which characterizes the \textit{genuine} type-I class.
The insets of Fig.~\ref{fig:fig1}
depict the magnetization curve of two selected $\kappa$ values belonging to the two classes, chosen as $0.28$ and $0.33$ for $L=16\lambda$.
The left inset  depicts a magnetization that goes directly from the normal ($M=0$) to the Meissner state,
while the right inset shows a spike highlighted by the black ellipse that corresponds to the coalescence of flux in form of vortices and their subsequent exit at an infinitesimally lower field.
The magnetization curves show the up (blue line)
and down (red line) branches for both insets.

%----------------------------------------------------------------
\begin{figure}
\centering
\includegraphics[width=0.48\textwidth]{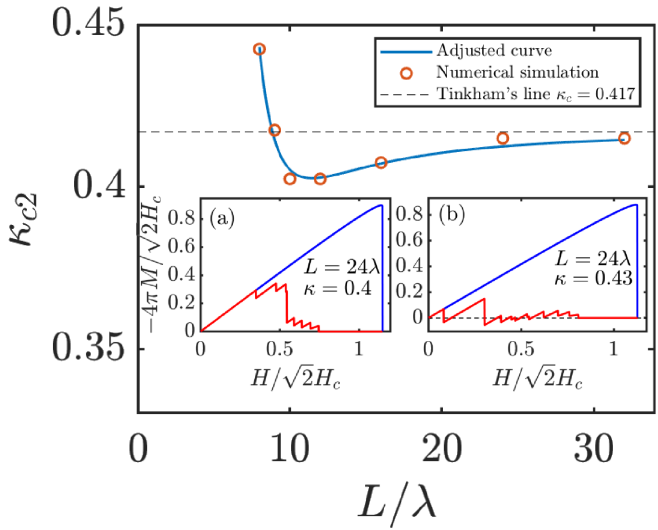}
\caption{(Color online) The magnetization in descending field reveals the transition at $\kappa_{c2}$, shown in the main panel as a function of $L$,  from paramagnetic to diamagnetic response. Insets (a) and (b) display typical magnetization curves below and above the transition.
    Inset (b) shows a still  paramagnetic magnetization while in panel (a) it has become totally diamagnetic.}
\label{fig:fig2}
\end{figure}
%----------------------------------------------------------------

Fig.~\ref{fig:fig2}  features  the \textit{dia-para} transition. The main panel presents $\kappa_{c2}$ as a function of $L$. The small red circles indicates the numerically obtained $\kappa_{c2}$ values of $0.4425$,  $0.4175$,  $0.4025$, $0.4025$, $0.4075$,  $0.415$, and  $0.415$, for $L/\lambda$ equal to $8$, $9$, $10$, $12$, $16$, $24$, and  $32$, respectively.
Although $\kappa_{c2}$ is a non-monotonic function of $L$, it  asymptotically approaches a limiting value for large $L$, suggestively close to  $\kappa_c$, which is the value that delimits the \textit{genuine-intermediate} transition in the macroscopic limit.
Insets (a) and (b) of Fig.~\ref{fig:fig2} depict the transition occurring at $\kappa_{c2}$ for the case of
$L=24\lambda$ through two selected values of $\kappa$, each  characterizing one side of the transition.
Inset (a) shows the typical diamagnetic behavior for $\kappa=0.4<\kappa_{c2}$, the magnetization  is  always negative for any value of the applied field.
Inset (b) presents the magnetization for the case with $\kappa=0.43>\kappa_{c2}$ at which paramagnetic regions exists in the down branch and alternates with diamagnetic ones, thus not qualifying as a totally diamagnetic response.

%----------------------------------------------------------------
\begin{figure}
    \centering
    \includegraphics[width=0.48\textwidth]{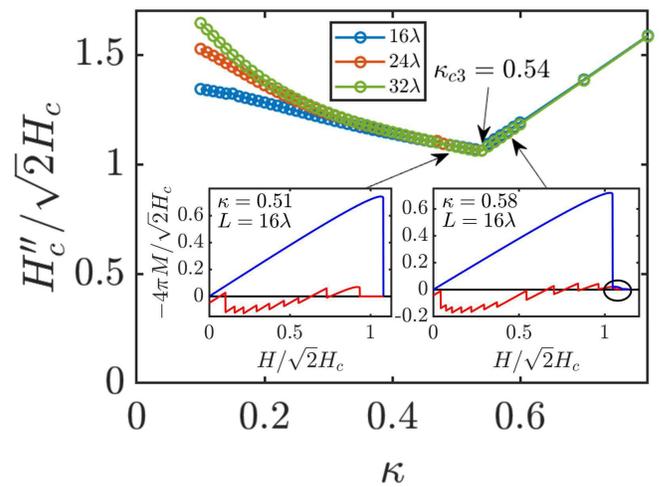}
    \caption{(Color online) The magnetization is a maximum at the field $H^{\prime}_c$ (the peak of $-M$) and vanishes at the field $H^{\prime\prime}_c$ where superconductivity is destroyed and the normal state is restored. $H^{\prime\prime}_c$ versus $\kappa$ is shown here and clearly points to a $\kappa_{c3}$ transition.
    The insets show typical magnetization curves below and above this transition.
    The left inset shows the situation that $H^{\prime\prime}_c \approx H^{\prime}_c$ and no vortices
    are present whereas the right
    inset illustrates the situation  $H^{\prime\prime}_c > H^{\prime}_c$ in which vortices do nucleate in the region highlighted by an ellipse.}
    \label{fig:fig3}
\end{figure}
%----------------------------------------------------------------
Fig.~\ref{fig:fig3} shows the  \textit{quasi} type-II transition, signaled in the ascending magnetization by two notable fields, namely, $H^{\prime\prime}_c$ and $H^{\prime}_c$.
For $\kappa< \kappa_{c3}$,  the maximum of the magnetization is immediately followed by its sudden drop to zero, $H^{\prime\prime}_c \approx H^{\prime}_c$. However for $\kappa > \kappa_{c3}$ the two fields depart from each other and  for increasing $\kappa$, $H^{\prime\prime}_c- H^{\prime}_c$ also increases and vortices are observed in the superconductor.
The kink in the curve $H^{\prime\prime}_c$ versus $\kappa$ shown in Fig.~\ref{fig:fig3} defines $\kappa_{c3}$.
The insets of Fig.~\ref{fig:fig3} display situations below (left) and above (right) the transition.
The left one shows the magnetization curve going from the Meissner state directly to the normal state, whereas the right one, in contrast, presents a vortex state in between the Meissner and normal states. The black ellipse in this inset highlights the vortex state region.
Remarkably, this transition is found to occur at $\kappa_{c3}=0.54$ for any size $L$. Interestingly, it is found that, apart from small numerical deviations, $H^{\prime\prime}_c \approx H^{\prime}_{c3}$ for  $\kappa > \kappa_{c3}$.
Concerning the Gibbs free energy of the  \textit{quasi} type-II class, it is negative though very close to zero. This small negative value is still sufficient to render it  slightly below the normal state energy, which is zero (see the supplementary material). This  vortex regime is subsequent to the peak of the magnetization, which lies in an energetically metastable regime. This is in contrast with the standard type-II superconductor, where the peak of the magnetization is within a totally stable regime. We find that $\kappa=0.8$ is still \textit{quasi} type-II behavior but not specify the upper boundary which ought to be connected to the stability of the magnetization peak.

%----------------------------------------------------------------
\begin{figure}
    \centering
    \includegraphics[width=0.48\textwidth]{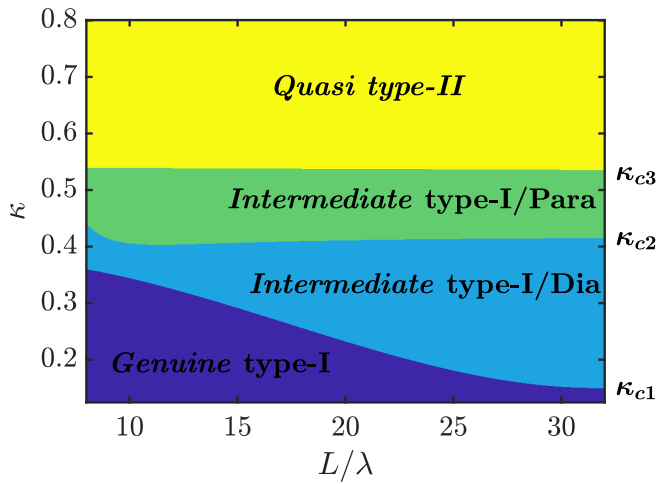}
    \caption{(Color online) $\kappa-L$ phase diagram for the system under study in the ranges $0.125<\kappa<0.8$ and $8\lambda<L<32\lambda$. As given in the figure, the dark blue, the light blue, the green and the yellow regions corresponds to \textit{Genuine} type-I, \textit{Intermediate} type-I with diamagnetism, \textit{Intermediate} type-I with the occurrence paragmetism in the down branch, \textit{quasi} type-II, respectively.}
    \label{fig:fig4}
\end{figure}
%----------------------------------------------------------------

Fig.~\ref{fig:fig4} displays the $\kappa$ versus $L$ phase diagram containing all the transitions discussed here.
The GL theory ceases to be valid at  $L = \xi$, and for this reason the diagram features $\kappa \ge 0.125$, which guarantees $L > \xi$ for all $L$ considered.
The delimiting curves separating any two regions are obtained by a fitting process.
The $\kappa_{ci}(L)$ lines are indicated at the right margin of the figure and they separate the four regions, namely \textit{Genuine} and \textit{Intermediate}, the latter  split into sub classes known as  dia, para, and \textit{quasi} type-II.
Hence  Fig.~\ref{fig:fig4} is the generalization for the  mesoscopic
superconductors of the de-Gennes-Tinkham transition found at $\kappa_c$ for the macroscopic superconductor.
\\

\section{Conclusions}
In summary, we show  that mesoscopic type-I superconductors have intrinsic transitions in $\kappa$. The \textit{genuine} type-I behavior is only possible below $\kappa_{c1}$, and above it, vortices exist in this so-called \textit{intermediate} type-I class that has a rich structure with a transition from paramagnetic to diamagnetic response, in descending field ($\kappa_{c2}$), and a \textit{quasi} type-II behavior, in ascending field ($\kappa_{c3}$).
\\

\begin{acknowledgments}
LRC thanks the Brazilian Agency Funda\c c\~ao de Amparo à Pesquisa do Estado de S\~ao Paulo (FAPESP) for financial support (process number 20/03947-2).
ARdeCR, MMD and ES thank FACEPE and CAPES for financial support with the project number APQ-0198-1.05/14.
ES thanks the Brazilan Agency Funda\c c\~ao de Amparo à Pesquisa do Estado de S\~ao Paulo (FAPESP) for financial support
(process number 12/04388-0).
We also gratefully acknowledge the support of the NVIDIA Corporation with the donation of GPUs for our research.
\end{acknowledgments}

%\nocite{*}

%\bibliography{references}% Produces the bibliography via BibTeX.
\bibliography{references}% Produces the bibliography via BibTeX.

\begin{thebibliography}{47}%
\makeatletter
\providecommand \@ifxundefined [1]{%
 \@ifx{#1\undefined}
}%
\providecommand \@ifnum [1]{%
 \ifnum #1\expandafter \@firstoftwo
 \else \expandafter \@secondoftwo
 \fi
}%
\providecommand \@ifx [1]{%
 \ifx #1\expandafter \@firstoftwo
 \else \expandafter \@secondoftwo
 \fi
}%
\providecommand \natexlab [1]{#1}%
\providecommand \enquote  [1]{``#1''}%
\providecommand \bibnamefont  [1]{#1}%
\providecommand \bibfnamefont [1]{#1}%
\providecommand \citenamefont [1]{#1}%
\providecommand \href@noop [0]{\@secondoftwo}%
\providecommand \href [0]{\begingroup \@sanitize@url \@href}%
\providecommand \@href[1]{\@@startlink{#1}\@@href}%
\providecommand \@@href[1]{\endgroup#1\@@endlink}%
\providecommand \@sanitize@url [0]{\catcode `\\12\catcode `\$12\catcode
  `\&12\catcode `\#12\catcode `\^12\catcode `\_12\catcode `\%12\relax}%
\providecommand \@@startlink[1]{}%
\providecommand \@@endlink[0]{}%
\providecommand \url  [0]{\begingroup\@sanitize@url \@url }%
\providecommand \@url [1]{\endgroup\@href {#1}{\urlprefix }}%
\providecommand \urlprefix  [0]{URL }%
\providecommand \Eprint [0]{\href }%
\providecommand \doibase [0]{https://doi.org/}%
\providecommand \selectlanguage [0]{\@gobble}%
\providecommand \bibinfo  [0]{\@secondoftwo}%
\providecommand \bibfield  [0]{\@secondoftwo}%
\providecommand \translation [1]{[#1]}%
\providecommand \BibitemOpen [0]{}%
\providecommand \bibitemStop [0]{}%
\providecommand \bibitemNoStop [0]{.\EOS\space}%
\providecommand \EOS [0]{\spacefactor3000\relax}%
\providecommand \BibitemShut  [1]{\csname bibitem#1\endcsname}%
\let\auto@bib@innerbib\@empty
%</preamble>
\bibitem [{\citenamefont {Tinkham}(2004)}]{tinkham04}%
  \BibitemOpen
  \bibfield  {author} {\bibinfo {author} {\bibfnamefont {M.}~\bibnamefont
  {Tinkham}},\ }\href {http://www.worldcat.org/isbn/0486435032} {\emph
  {\bibinfo {title} {Introduction to Superconductivity}}},\ \bibinfo {edition}
  {2nd}\ ed.\ (\bibinfo  {publisher} {Dover Publications},\ \bibinfo {year}
  {2004})\ Chap.~\bibinfo {chapter} {4}, pp.\ \bibinfo {pages}
  {135--138}\BibitemShut {NoStop}%
\bibitem [{\citenamefont {de~Gennes}(1966)}]{degennes66}%
  \BibitemOpen
  \bibfield  {author} {\bibinfo {author} {\bibfnamefont {P.~G.}\ \bibnamefont
  {de~Gennes}},\ }\href {https://books.google.com.br/books?id=M8A8AAAAIAAJ}
  {\emph {\bibinfo {title} {Superconductivity of metals and alloys}}}\
  (\bibinfo  {publisher} {W. A. Benjamin},\ \bibinfo {year} {1966})\
  Chap.~\bibinfo {chapter} {6}, pp.\ \bibinfo {pages} {199--201}\BibitemShut
  {NoStop}%
\bibitem [{\citenamefont {Shepelev}(2010)}]{shepelev10}%
  \BibitemOpen
  \bibfield  {author} {\bibinfo {author} {\bibfnamefont {A.}~\bibnamefont
  {Shepelev}},\ }\bibfield  {title} {\bibinfo {title} {The discovery of type
  {II} superconductors (shubnikov phase)},\ }in\ \href
  {https://doi.org/10.5772/10117} {\emph {\bibinfo {booktitle}
  {Superconductor}}},\ \bibinfo {editor} {edited by\ \bibinfo {editor}
  {\bibfnamefont {A.~M.}\ \bibnamefont {Luiz}}}\ (\bibinfo  {publisher}
  {IntechOpen},\ \bibinfo {address} {Rijeka},\ \bibinfo {year} {2010})\
  Chap.~\bibinfo {chapter} {2}\BibitemShut {NoStop}%
\bibitem [{\citenamefont {Abrikosov}(1957)}]{abrikosov56}%
  \BibitemOpen
  \bibfield  {author} {\bibinfo {author} {\bibfnamefont {A.~A.}\ \bibnamefont
  {Abrikosov}},\ }\bibfield  {title} {\bibinfo {title} {{On the Magnetic
  properties of superconductors of the second group}},\ }\href
  {http://www.jetp.ac.ru/cgi-bin/dn/e_005_06_1174.pdf} {\bibfield  {journal}
  {\bibinfo  {journal} {Sov. Phys. JETP}\ }\textbf {\bibinfo {volume} {5}},\
  \bibinfo {pages} {1174} (\bibinfo {year} {1957})},\ \bibinfo {note} {[Zh.
  Eksp. Teor. Fiz.32,1442(1957)]}\BibitemShut {NoStop}%
%%CITATION = SPHJA,5,1174;%%
\bibitem [{\citenamefont {Nielsen}\ and\ \citenamefont
  {Olesen}(1973)}]{nielsen73}%
  \BibitemOpen
  \bibfield  {author} {\bibinfo {author} {\bibfnamefont {H.}~\bibnamefont
  {Nielsen}}\ and\ \bibinfo {author} {\bibfnamefont {P.}~\bibnamefont
  {Olesen}},\ }\bibfield  {title} {\bibinfo {title} {Vortex-line models for
  dual strings},\ }\href
  {https://doi.org/https://doi.org/10.1016/0550-3213(73)90350-7} {\bibfield
  {journal} {\bibinfo  {journal} {Nucl. Phys. B}\ }\textbf {\bibinfo {volume}
  {61}},\ \bibinfo {pages} {45 } (\bibinfo {year} {1973})}\BibitemShut
  {NoStop}%
\bibitem [{\citenamefont {Bogomolny}(1976)}]{bogomolny76}%
  \BibitemOpen
  \bibfield  {author} {\bibinfo {author} {\bibfnamefont {E.~B.}\ \bibnamefont
  {Bogomolny}},\ }\bibfield  {title} {\bibinfo {title} {The stability of
  classical solutions},\ }\href
  {https://www.docsity.com/pt/the-stability-of-classical-solutions/4895287/}
  {\bibfield  {journal} {\bibinfo  {journal} {Sov. J. Nucl. Phys.}\ }\textbf
  {\bibinfo {volume} {24}},\ \bibinfo {pages} {449} (\bibinfo {year}
  {1976})}\BibitemShut {NoStop}%
\bibitem [{\citenamefont {Luk'yanchuk}(2001)}]{lukyanchuk01}%
  \BibitemOpen
  \bibfield  {author} {\bibinfo {author} {\bibfnamefont {I.}~\bibnamefont
  {Luk'yanchuk}},\ }\bibfield  {title} {\bibinfo {title} {Theory of
  superconductors with $\ensuremath{\kappa}$ close to $1/\sqrt{2}$},\ }\href
  {https://doi.org/10.1103/PhysRevB.63.174504} {\bibfield  {journal} {\bibinfo
  {journal} {Phys. Rev. B}\ }\textbf {\bibinfo {volume} {63}},\ \bibinfo
  {pages} {174504} (\bibinfo {year} {2001})}\BibitemShut {NoStop}%
\bibitem [{\citenamefont {Webb}\ \emph {et~al.}(2015)\citenamefont {Webb},
  \citenamefont {Marsiglio},\ and\ \citenamefont {Hirsch}}]{webb15}%
  \BibitemOpen
  \bibfield  {author} {\bibinfo {author} {\bibfnamefont {G.}~\bibnamefont
  {Webb}}, \bibinfo {author} {\bibfnamefont {F.}~\bibnamefont {Marsiglio}},\
  and\ \bibinfo {author} {\bibfnamefont {J.}~\bibnamefont {Hirsch}},\
  }\bibfield  {title} {\bibinfo {title} {Superconductivity in the elements,
  alloys and simple compounds},\ }\href
  {https://doi.org/https://doi.org/10.1016/j.physc.2015.02.037} {\bibfield
  {journal} {\bibinfo  {journal} {Physica C}\ }\textbf {\bibinfo {volume}
  {514}},\ \bibinfo {pages} {17 } (\bibinfo {year} {2015})}\BibitemShut
  {NoStop}%
\bibitem [{\citenamefont {Wesche}(2017)}]{wesche17}%
  \BibitemOpen
  \bibfield  {author} {\bibinfo {author} {\bibfnamefont {R.}~\bibnamefont
  {Wesche}},\ }\bibinfo {title} {High-temperature superconductors}\ (\bibinfo
  {publisher} {Springer International Publishing},\ \bibinfo {address} {Cham},\
  \bibinfo {year} {2017})\ p.~\bibinfo {pages} {1}\BibitemShut {NoStop}%
\bibitem [{\citenamefont {Gottlieb}\ \emph {et~al.}(1992)\citenamefont
  {Gottlieb}, \citenamefont {Lasjaunias}, \citenamefont {Tholence},
  \citenamefont {Laborde}, \citenamefont {Thomas},\ and\ \citenamefont
  {Madar}}]{gottlieb92}%
  \BibitemOpen
  \bibfield  {author} {\bibinfo {author} {\bibfnamefont {U.}~\bibnamefont
  {Gottlieb}}, \bibinfo {author} {\bibfnamefont {J.~C.}\ \bibnamefont
  {Lasjaunias}}, \bibinfo {author} {\bibfnamefont {J.~L.}\ \bibnamefont
  {Tholence}}, \bibinfo {author} {\bibfnamefont {O.}~\bibnamefont {Laborde}},
  \bibinfo {author} {\bibfnamefont {O.}~\bibnamefont {Thomas}},\ and\ \bibinfo
  {author} {\bibfnamefont {R.}~\bibnamefont {Madar}},\ }\bibfield  {title}
  {\bibinfo {title} {Superconductivity in ${\mathrm{tasi}}_{2}$ single
  crystals},\ }\href {https://doi.org/10.1103/PhysRevB.45.4803} {\bibfield
  {journal} {\bibinfo  {journal} {Phys. Rev. B}\ }\textbf {\bibinfo {volume}
  {45}},\ \bibinfo {pages} {4803} (\bibinfo {year} {1992})}\BibitemShut
  {NoStop}%
\bibitem [{\citenamefont {Kriener}\ \emph {et~al.}(2008)\citenamefont
  {Kriener}, \citenamefont {Muranaka}, \citenamefont {Kato}, \citenamefont
  {Ren}, \citenamefont {Akimitsu},\ and\ \citenamefont {Maeno}}]{kriener08}%
  \BibitemOpen
  \bibfield  {author} {\bibinfo {author} {\bibfnamefont {M.}~\bibnamefont
  {Kriener}}, \bibinfo {author} {\bibfnamefont {T.}~\bibnamefont {Muranaka}},
  \bibinfo {author} {\bibfnamefont {J.}~\bibnamefont {Kato}}, \bibinfo {author}
  {\bibfnamefont {Z.-A.}\ \bibnamefont {Ren}}, \bibinfo {author} {\bibfnamefont
  {J.}~\bibnamefont {Akimitsu}},\ and\ \bibinfo {author} {\bibfnamefont
  {Y.}~\bibnamefont {Maeno}},\ }\bibfield  {title} {\bibinfo {title}
  {Superconductivity in heavily boron-doped silicon carbide},\ }\href
  {https://doi.org/10.1088/1468-6996/9/4/044205} {\bibfield  {journal}
  {\bibinfo  {journal} {Sci. Technol. Adv. Mater.}\ }\textbf {\bibinfo {volume}
  {9}},\ \bibinfo {pages} {044205} (\bibinfo {year} {2008})}\BibitemShut
  {NoStop}%
\bibitem [{\citenamefont {Zhao}\ \emph {et~al.}(2012)\citenamefont {Zhao},
  \citenamefont {Lausberg}, \citenamefont {Kim}, \citenamefont {Tanatar},
  \citenamefont {Brando}, \citenamefont {Prozorov},\ and\ \citenamefont
  {Morosan}}]{zhao12}%
  \BibitemOpen
  \bibfield  {author} {\bibinfo {author} {\bibfnamefont {L.~L.}\ \bibnamefont
  {Zhao}}, \bibinfo {author} {\bibfnamefont {S.}~\bibnamefont {Lausberg}},
  \bibinfo {author} {\bibfnamefont {H.}~\bibnamefont {Kim}}, \bibinfo {author}
  {\bibfnamefont {M.~A.}\ \bibnamefont {Tanatar}}, \bibinfo {author}
  {\bibfnamefont {M.}~\bibnamefont {Brando}}, \bibinfo {author} {\bibfnamefont
  {R.}~\bibnamefont {Prozorov}},\ and\ \bibinfo {author} {\bibfnamefont
  {E.}~\bibnamefont {Morosan}},\ }\bibfield  {title} {\bibinfo {title}
  {Type-{I} superconductivity in ybsb${}_{2}$ single crystals},\ }\href
  {https://doi.org/10.1103/PhysRevB.85.214526} {\bibfield  {journal} {\bibinfo
  {journal} {Phys. Rev. B}\ }\textbf {\bibinfo {volume} {85}},\ \bibinfo
  {pages} {214526} (\bibinfo {year} {2012})}\BibitemShut {NoStop}%
\bibitem [{\citenamefont {Arantes}\ \emph {et~al.}(2019)\citenamefont
  {Arantes}, \citenamefont {Aristiz\'abal-Giraldo} \emph {et~al.}}]{avila19}%
  \BibitemOpen
  \bibfield  {author} {\bibinfo {author} {\bibfnamefont {F.~R.}\ \bibnamefont
  {Arantes}}, \bibinfo {author} {\bibnamefont {Aristiz\'abal-Giraldo}}, \emph
  {et~al.},\ }\bibfield  {title} {\bibinfo {title} {Superconductivity in
  monocrystalline ${\mathrm{ynisi}}_{3}$ and ${\mathrm{lunisi}}_{3}$},\ }\href
  {https://doi.org/10.1103/PhysRevB.99.224505} {\bibfield  {journal} {\bibinfo
  {journal} {Phys. Rev. B}\ }\textbf {\bibinfo {volume} {99}},\ \bibinfo
  {pages} {224505} (\bibinfo {year} {2019})}\BibitemShut {NoStop}%
\bibitem [{\citenamefont {Feder}(1967)}]{feder67}%
  \BibitemOpen
  \bibfield  {author} {\bibinfo {author} {\bibfnamefont {J.}~\bibnamefont
  {Feder}},\ }\bibfield  {title} {\bibinfo {title} {Comments of the
  supercooling field for superconductors with k values near 0.4},\ }\href
  {https://doi.org/https://doi.org/10.1016/0038-1098(67)90277-3} {\bibfield
  {journal} {\bibinfo  {journal} {Solid State Commun.}\ }\textbf {\bibinfo
  {volume} {5}},\ \bibinfo {pages} {299 } (\bibinfo {year} {1967})}\BibitemShut
  {NoStop}%
\bibitem [{\citenamefont {Park}(1967)}]{park67}%
  \BibitemOpen
  \bibfield  {author} {\bibinfo {author} {\bibfnamefont {J.}~\bibnamefont
  {Park}},\ }\bibfield  {title} {\bibinfo {title} {Metastable states of the
  superconducting surface sheath in decreasing fields},\ }\href
  {https://doi.org/https://doi.org/10.1016/0038-1098(67)90084-1} {\bibfield
  {journal} {\bibinfo  {journal} {Solid State Commun.}\ }\textbf {\bibinfo
  {volume} {5}},\ \bibinfo {pages} {645 } (\bibinfo {year} {1967})}\BibitemShut
  {NoStop}%
\bibitem [{\citenamefont {Buchanan}\ \emph {et~al.}(1965)\citenamefont
  {Buchanan}, \citenamefont {Chang},\ and\ \citenamefont {Serin}}]{buchanan65}%
  \BibitemOpen
  \bibfield  {author} {\bibinfo {author} {\bibfnamefont {J.}~\bibnamefont
  {Buchanan}}, \bibinfo {author} {\bibfnamefont {G.}~\bibnamefont {Chang}},\
  and\ \bibinfo {author} {\bibfnamefont {B.}~\bibnamefont {Serin}},\ }\bibfield
   {title} {\bibinfo {title} {The ginzburg-landau parameter of tantalum},\
  }\href {https://doi.org/https://doi.org/10.1016/0022-3697(65)90016-8}
  {\bibfield  {journal} {\bibinfo  {journal} {J. Phys. Chem. Solids}\ }\textbf
  {\bibinfo {volume} {26}},\ \bibinfo {pages} {1183 } (\bibinfo {year}
  {1965})}\BibitemShut {NoStop}%
\bibitem [{\citenamefont {McEvoy}\ \emph {et~al.}(1967)\citenamefont {McEvoy},
  \citenamefont {Jones},\ and\ \citenamefont {Park}}]{mcevoy67}%
  \BibitemOpen
  \bibfield  {author} {\bibinfo {author} {\bibfnamefont {J.}~\bibnamefont
  {McEvoy}}, \bibinfo {author} {\bibfnamefont {D.}~\bibnamefont {Jones}},\ and\
  \bibinfo {author} {\bibfnamefont {J.}~\bibnamefont {Park}},\ }\bibfield
  {title} {\bibinfo {title} {Supercooling of superconductors below the surface
  nucleation field},\ }\href
  {https://doi.org/https://doi.org/10.1016/0038-1098(67)90083-X} {\bibfield
  {journal} {\bibinfo  {journal} {Solid State Commun.}\ }\textbf {\bibinfo
  {volume} {5}},\ \bibinfo {pages} {641 } (\bibinfo {year} {1967})}\BibitemShut
  {NoStop}%
\bibitem [{\citenamefont {Maloney}\ \emph {et~al.}(1972)\citenamefont
  {Maloney}, \citenamefont {de~la Cruz},\ and\ \citenamefont
  {Cardona}}]{paco72}%
  \BibitemOpen
  \bibfield  {author} {\bibinfo {author} {\bibfnamefont {M.~D.}\ \bibnamefont
  {Maloney}}, \bibinfo {author} {\bibfnamefont {F.}~\bibnamefont {de~la
  Cruz}},\ and\ \bibinfo {author} {\bibfnamefont {M.}~\bibnamefont {Cardona}},\
  }\bibfield  {title} {\bibinfo {title} {Superconducting parameters and size
  effects of aluminum films and foils},\ }\href
  {https://doi.org/10.1103/PhysRevB.5.3558} {\bibfield  {journal} {\bibinfo
  {journal} {Phys. Rev. B}\ }\textbf {\bibinfo {volume} {5}},\ \bibinfo {pages}
  {3558} (\bibinfo {year} {1972})}\BibitemShut {NoStop}%
\bibitem [{\citenamefont {Blot}\ \emph {et~al.}(1978)\citenamefont {Blot},
  \citenamefont {Pellan},\ and\ \citenamefont {Rosenblatt}}]{blot78}%
  \BibitemOpen
  \bibfield  {author} {\bibinfo {author} {\bibfnamefont {J.}~\bibnamefont
  {Blot}}, \bibinfo {author} {\bibfnamefont {Y.}~\bibnamefont {Pellan}},\ and\
  \bibinfo {author} {\bibfnamefont {J.}~\bibnamefont {Rosenblatt}},\ }\bibfield
   {title} {\bibinfo {title} {Metastable states of superconducting indium
  films},\ }\href {https://doi.org/10.1007/BF00116205} {\bibfield  {journal}
  {\bibinfo  {journal} {J. Low Temp. Phys.}\ }\textbf {\bibinfo {volume}
  {30}},\ \bibinfo {pages} {669} (\bibinfo {year} {1978})}\BibitemShut
  {NoStop}%
\bibitem [{\citenamefont {Zharkov}(2005)}]{zharkov05}%
  \BibitemOpen
  \bibfield  {author} {\bibinfo {author} {\bibfnamefont {G.~F.}\ \bibnamefont
  {Zharkov}},\ }\bibfield  {title} {\bibinfo {title} {Superconducting states of
  the cylinder with a single vortex in magnetic field according to the
  ginzburg-landau theory},\ }\href {https://doi.org/10.2478/BF02476508}
  {\bibfield  {journal} {\bibinfo  {journal} {Centr. Eur. J. Phys.}\ }\textbf
  {\bibinfo {volume} {3}},\ \bibinfo {pages} {77} (\bibinfo {year}
  {2005})}\BibitemShut {NoStop}%
\bibitem [{\citenamefont {Saint-James}\ and\ \citenamefont
  {de~Gennes}(1963)}]{saintjames63}%
  \BibitemOpen
  \bibfield  {author} {\bibinfo {author} {\bibfnamefont {D.}~\bibnamefont
  {Saint-James}}\ and\ \bibinfo {author} {\bibfnamefont {P.~G.}\ \bibnamefont
  {de~Gennes}},\ }\bibfield  {title} {\bibinfo {title} {Onset of
  superconductivity in decreasing fields},\ }\href
  {https://doi.org/https://doi.org/10.1016/0031-9163(63)90047-7} {\bibfield
  {journal} {\bibinfo  {journal} {Phys. Lett.}\ }\textbf {\bibinfo {volume}
  {7}},\ \bibinfo {pages} {306} (\bibinfo {year} {1963})}\BibitemShut {NoStop}%
\bibitem [{\citenamefont {Saint-James}(1965)}]{saintjames65}%
  \BibitemOpen
  \bibfield  {author} {\bibinfo {author} {\bibfnamefont {D.}~\bibnamefont
  {Saint-James}},\ }\bibfield  {title} {\bibinfo {title} {Etude du champ
  critique hc3 dans une geometrie cylindrique},\ }\href
  {https://doi.org/https://doi.org/10.1016/0031-9163(65)91101-7} {\bibfield
  {journal} {\bibinfo  {journal} {Phys. Lett.}\ }\textbf {\bibinfo {volume}
  {15}},\ \bibinfo {pages} {13} (\bibinfo {year} {1965})}\BibitemShut {NoStop}%
\bibitem [{\citenamefont {Kwasnitza}\ and\ \citenamefont
  {Rupp}(1966)}]{kwasnitza66}%
  \BibitemOpen
  \bibfield  {author} {\bibinfo {author} {\bibfnamefont {K.}~\bibnamefont
  {Kwasnitza}}\ and\ \bibinfo {author} {\bibfnamefont {G.}~\bibnamefont
  {Rupp}},\ }\bibfield  {title} {\bibinfo {title} {Measurement of critical
  field hc3 and critical surface current in superconducting v-ti alloys up to
  30 koe},\ }\href
  {https://doi.org/https://doi.org/10.1016/0031-9163(66)90247-2} {\bibfield
  {journal} {\bibinfo  {journal} {Phys. Lett.}\ }\textbf {\bibinfo {volume}
  {23}},\ \bibinfo {pages} {40} (\bibinfo {year} {1966})}\BibitemShut {NoStop}%
\bibitem [{\citenamefont {Hauser}\ \emph {et~al.}(1974)\citenamefont {Hauser},
  \citenamefont {Wang},\ and\ \citenamefont {Kittel}}]{hauser74}%
  \BibitemOpen
  \bibfield  {author} {\bibinfo {author} {\bibfnamefont {J.}~\bibnamefont
  {Hauser}}, \bibinfo {author} {\bibfnamefont {J.-Y.}\ \bibnamefont {Wang}},\
  and\ \bibinfo {author} {\bibfnamefont {C.}~\bibnamefont {Kittel}},\
  }\bibfield  {title} {\bibinfo {title} {Calculation of the surface critical
  field hc3 for a cylindrical cavity},\ }\href
  {https://doi.org/https://doi.org/10.1016/0375-9601(74)90094-2} {\bibfield
  {journal} {\bibinfo  {journal} {Phys. Lett. A}\ }\textbf {\bibinfo {volume}
  {47}},\ \bibinfo {pages} {34} (\bibinfo {year} {1974})}\BibitemShut {NoStop}%
\bibitem [{\citenamefont {Askerzade}(2003)}]{askerzade03}%
  \BibitemOpen
  \bibfield  {author} {\bibinfo {author} {\bibfnamefont {I.~N.}\ \bibnamefont
  {Askerzade}},\ }\bibfield  {title} {\bibinfo {title} {Order parameter
  anisotropy of mgb2 using specific heat jump of layered superconductors},\
  }\href {https://www.ias.ac.in/article/fulltext/pram/061/06/1145-1149}
  {\bibfield  {journal} {\bibinfo  {journal} {Pramana}\ }\textbf {\bibinfo
  {volume} {61}},\ \bibinfo {pages} {1145} (\bibinfo {year}
  {2003})}\BibitemShut {NoStop}%
\bibitem [{\citenamefont {Changjan}\ \emph {et~al.}(2017)\citenamefont
  {Changjan}, \citenamefont {Meakniti},\ and\ \citenamefont
  {Udomsamuthirun}}]{changjan17}%
  \BibitemOpen
  \bibfield  {author} {\bibinfo {author} {\bibfnamefont {A.}~\bibnamefont
  {Changjan}}, \bibinfo {author} {\bibfnamefont {S.}~\bibnamefont {Meakniti}},\
  and\ \bibinfo {author} {\bibfnamefont {P.}~\bibnamefont {Udomsamuthirun}},\
  }\bibfield  {title} {\bibinfo {title} {The temperature-dependent surface
  critical magnetic field (hc3) of magnetic superconductors: Applied to lead
  bismuth (pb82bi18) superconductors},\ }\href
  {https://doi.org/https://doi.org/10.1016/j.jpcs.2017.03.022} {\bibfield
  {journal} {\bibinfo  {journal} {J. Phys. Chem. Solids}\ }\textbf {\bibinfo
  {volume} {107}},\ \bibinfo {pages} {32 } (\bibinfo {year}
  {2017})}\BibitemShut {NoStop}%
\bibitem [{\citenamefont {Xie}\ \emph {et~al.}(2017)\citenamefont {Xie},
  \citenamefont {Kogan}, \citenamefont {Khodas},\ and\ \citenamefont
  {Levchenko}}]{xie17}%
  \BibitemOpen
  \bibfield  {author} {\bibinfo {author} {\bibfnamefont {H.-Y.}\ \bibnamefont
  {Xie}}, \bibinfo {author} {\bibfnamefont {V.~G.}\ \bibnamefont {Kogan}},
  \bibinfo {author} {\bibfnamefont {M.}~\bibnamefont {Khodas}},\ and\ \bibinfo
  {author} {\bibfnamefont {A.}~\bibnamefont {Levchenko}},\ }\bibfield  {title}
  {\bibinfo {title} {Onset of surface superconductivity beyond the
  saint-james--de gennes limit},\ }\href
  {https://doi.org/10.1103/PhysRevB.96.104516} {\bibfield  {journal} {\bibinfo
  {journal} {Phys. Rev. B}\ }\textbf {\bibinfo {volume} {96}},\ \bibinfo
  {pages} {104516} (\bibinfo {year} {2017})}\BibitemShut {NoStop}%
\bibitem [{\citenamefont {Christiansen}\ and\ \citenamefont
  {Smith}(1968)}]{christiansen68}%
  \BibitemOpen
  \bibfield  {author} {\bibinfo {author} {\bibfnamefont {P.~V.}\ \bibnamefont
  {Christiansen}}\ and\ \bibinfo {author} {\bibfnamefont {H.}~\bibnamefont
  {Smith}},\ }\bibfield  {title} {\bibinfo {title} {Ginzburg-landau theory of
  surface superconductivity and magnetic hysteresis},\ }\href
  {https://doi.org/10.1103/PhysRev.171.445} {\bibfield  {journal} {\bibinfo
  {journal} {Phys. Rev.}\ }\textbf {\bibinfo {volume} {171}},\ \bibinfo {pages}
  {445} (\bibinfo {year} {1968})}\BibitemShut {NoStop}%
\bibitem [{\citenamefont {Jung}\ \emph {et~al.}(2002)\citenamefont {Jung},
  \citenamefont {Girard}, \citenamefont {Valko}, \citenamefont {Gomes},
  \citenamefont {Jeudy}, \citenamefont {Limagne},\ and\ \citenamefont
  {Waysand}}]{jung02}%
  \BibitemOpen
  \bibfield  {author} {\bibinfo {author} {\bibfnamefont {G.}~\bibnamefont
  {Jung}}, \bibinfo {author} {\bibfnamefont {T.}~\bibnamefont {Girard}},
  \bibinfo {author} {\bibfnamefont {P.}~\bibnamefont {Valko}}, \bibinfo
  {author} {\bibfnamefont {M.}~\bibnamefont {Gomes}}, \bibinfo {author}
  {\bibfnamefont {V.}~\bibnamefont {Jeudy}}, \bibinfo {author} {\bibfnamefont
  {D.}~\bibnamefont {Limagne}},\ and\ \bibinfo {author} {\bibfnamefont
  {G.}~\bibnamefont {Waysand}},\ }\bibfield  {title} {\bibinfo {title}
  {Expulsion of magnetic flux in a type-{I} superconducting strip},\ }\href
  {https://doi.org/https://doi.org/10.1016/S0921-4534(01)01126-1} {\bibfield
  {journal} {\bibinfo  {journal} {Physica C}\ }\textbf {\bibinfo {volume}
  {377}},\ \bibinfo {pages} {121 } (\bibinfo {year} {2002})}\BibitemShut
  {NoStop}%
\bibitem [{\citenamefont {Valko}\ \emph {et~al.}(2007)\citenamefont {Valko},
  \citenamefont {Gomes},\ and\ \citenamefont {Girard}}]{valko07}%
  \BibitemOpen
  \bibfield  {author} {\bibinfo {author} {\bibfnamefont {P.}~\bibnamefont
  {Valko}}, \bibinfo {author} {\bibfnamefont {M.~R.}\ \bibnamefont {Gomes}},\
  and\ \bibinfo {author} {\bibfnamefont {T.~A.}\ \bibnamefont {Girard}},\
  }\bibfield  {title} {\bibinfo {title} {Nucleation of superconductivity in
  thin type-{I} foils},\ }\href {https://doi.org/10.1103/PhysRevB.75.140504}
  {\bibfield  {journal} {\bibinfo  {journal} {Phys. Rev. B}\ }\textbf {\bibinfo
  {volume} {75}},\ \bibinfo {pages} {140504} (\bibinfo {year}
  {2007})}\BibitemShut {NoStop}%
\bibitem [{\citenamefont {Geim}\ \emph {et~al.}(1997)\citenamefont {Geim},
  \citenamefont {Grigorieva}, \citenamefont {Dubonos}, \citenamefont {Lok},
  \citenamefont {Maan}, \citenamefont {Filippov},\ and\ \citenamefont
  {Peeters}}]{geim97}%
  \BibitemOpen
  \bibfield  {author} {\bibinfo {author} {\bibfnamefont {A.~K.}\ \bibnamefont
  {Geim}}, \bibinfo {author} {\bibfnamefont {I.~V.}\ \bibnamefont
  {Grigorieva}}, \bibinfo {author} {\bibfnamefont {S.~V.}\ \bibnamefont
  {Dubonos}}, \bibinfo {author} {\bibfnamefont {J.~G.~S.}\ \bibnamefont {Lok}},
  \bibinfo {author} {\bibfnamefont {J.~C.}\ \bibnamefont {Maan}}, \bibinfo
  {author} {\bibfnamefont {A.~E.}\ \bibnamefont {Filippov}},\ and\ \bibinfo
  {author} {\bibfnamefont {F.~M.}\ \bibnamefont {Peeters}},\ }\bibfield
  {title} {\bibinfo {title} {Phase transitions in individual sub-micrometre
  superconductors},\ }\href {https://doi.org/10.1038/36797} {\bibfield
  {journal} {\bibinfo  {journal} {Nature}\ }\textbf {\bibinfo {volume} {390}},\
  \bibinfo {pages} {259} (\bibinfo {year} {1997})}\BibitemShut {NoStop}%
\bibitem [{\citenamefont {Berdiyorov}\ \emph {et~al.}(2009)\citenamefont
  {Berdiyorov}, \citenamefont {Hernandez},\ and\ \citenamefont
  {Peeters}}]{golib09}%
  \BibitemOpen
  \bibfield  {author} {\bibinfo {author} {\bibfnamefont {G.~R.}\ \bibnamefont
  {Berdiyorov}}, \bibinfo {author} {\bibfnamefont {A.~D.}\ \bibnamefont
  {Hernandez}},\ and\ \bibinfo {author} {\bibfnamefont {F.~M.}\ \bibnamefont
  {Peeters}},\ }\bibfield  {title} {\bibinfo {title} {Confinement effects on
  intermediate-state flux patterns in mesoscopic type-{I} superconductors},\
  }\href {https://doi.org/10.1103/PhysRevLett.103.267002} {\bibfield  {journal}
  {\bibinfo  {journal} {Phys. Rev. Lett.}\ }\textbf {\bibinfo {volume} {103}},\
  \bibinfo {pages} {267002} (\bibinfo {year} {2009})}\BibitemShut {NoStop}%
\bibitem [{\citenamefont {M\"uller}\ \emph {et~al.}(2012)\citenamefont
  {M\"uller}, \citenamefont {Milo{\v{s}}evi{\'c}}, \citenamefont {Dale},
  \citenamefont {Engbarth},\ and\ \citenamefont {Bending}}]{simon12}%
  \BibitemOpen
  \bibfield  {author} {\bibinfo {author} {\bibfnamefont {A.}~\bibnamefont
  {M\"uller}}, \bibinfo {author} {\bibfnamefont {M.}~\bibnamefont
  {Milo{\v{s}}evi{\'c}}}, \bibinfo {author} {\bibfnamefont {S.~E.~C.}\
  \bibnamefont {Dale}}, \bibinfo {author} {\bibfnamefont {M.~A.}\ \bibnamefont
  {Engbarth}},\ and\ \bibinfo {author} {\bibfnamefont {S.~J.}\ \bibnamefont
  {Bending}},\ }\bibfield  {title} {\bibinfo {title} {Magnetization
  measurements and ginzburg-landau simulations of micron-size
  $\ensuremath{\beta}$-tin samples: Evidence for an unusual critical behavior
  of mesoscopic type-{I} superconductors},\ }\href
  {https://doi.org/10.1103/PhysRevLett.109.197003} {\bibfield  {journal}
  {\bibinfo  {journal} {Phys. Rev. Lett.}\ }\textbf {\bibinfo {volume} {109}},\
  \bibinfo {pages} {197003} (\bibinfo {year} {2012})}\BibitemShut {NoStop}%
\bibitem [{\citenamefont {{de C. Romaguera}}\ \emph {et~al.}(2007)\citenamefont
  {{de C. Romaguera}}, \citenamefont {Doria},\ and\ \citenamefont
  {Peeters}}]{romaguera2007}%
  \BibitemOpen
  \bibfield  {author} {\bibinfo {author} {\bibfnamefont {A.~R.}\ \bibnamefont
  {{de C. Romaguera}}}, \bibinfo {author} {\bibfnamefont {M.~M.}\ \bibnamefont
  {Doria}},\ and\ \bibinfo {author} {\bibfnamefont {F.~M.}\ \bibnamefont
  {Peeters}},\ }\bibfield  {title} {\bibinfo {title} {Vortex pattern in a
  nanoscopic cylinder},\ }\href
  {https://doi.org/https://doi.org/10.1016/j.physc.2007.04.177} {\bibfield
  {journal} {\bibinfo  {journal} {Physica C}\ }\textbf {\bibinfo {volume}
  {460-462}},\ \bibinfo {pages} {1234 } (\bibinfo {year} {2007})}\BibitemShut
  {NoStop}%
\bibitem [{\citenamefont {Baelus}\ \emph {et~al.}(2006)\citenamefont {Baelus},
  \citenamefont {Kanda}, \citenamefont {Shimizu}, \citenamefont {Tadano},
  \citenamefont {Ootuka}, \citenamefont {Kadowaki},\ and\ \citenamefont
  {Peeters}}]{baelus2006}%
  \BibitemOpen
  \bibfield  {author} {\bibinfo {author} {\bibfnamefont {B.~J.}\ \bibnamefont
  {Baelus}}, \bibinfo {author} {\bibfnamefont {A.}~\bibnamefont {Kanda}},
  \bibinfo {author} {\bibfnamefont {N.}~\bibnamefont {Shimizu}}, \bibinfo
  {author} {\bibfnamefont {K.}~\bibnamefont {Tadano}}, \bibinfo {author}
  {\bibfnamefont {Y.}~\bibnamefont {Ootuka}}, \bibinfo {author} {\bibfnamefont
  {K.}~\bibnamefont {Kadowaki}},\ and\ \bibinfo {author} {\bibfnamefont
  {F.~M.}\ \bibnamefont {Peeters}},\ }\bibfield  {title} {\bibinfo {title}
  {Multivortex and giant vortex states near the expulsion and penetration
  fields in thin mesoscopic superconducting squares},\ }\href
  {https://doi.org/10.1103/PhysRevB.73.024514} {\bibfield  {journal} {\bibinfo
  {journal} {Phys. Rev. B}\ }\textbf {\bibinfo {volume} {73}},\ \bibinfo
  {pages} {024514} (\bibinfo {year} {2006})}\BibitemShut {NoStop}%
\bibitem [{\citenamefont {Shanenko}\ \emph {et~al.}(2011)\citenamefont
  {Shanenko}, \citenamefont {Milo\ifmmode \check{s}\else
  \v{s}\fi{}evi\ifmmode~\acute{c}\else \'{c}\fi{}}, \citenamefont {Peeters},\
  and\ \citenamefont {Vagov}}]{arkady11}%
  \BibitemOpen
  \bibfield  {author} {\bibinfo {author} {\bibfnamefont {A.~A.}\ \bibnamefont
  {Shanenko}}, \bibinfo {author} {\bibfnamefont {M.~V.}\ \bibnamefont
  {Milo\ifmmode \check{s}\else \v{s}\fi{}evi\ifmmode~\acute{c}\else
  \'{c}\fi{}}}, \bibinfo {author} {\bibfnamefont {F.~M.}\ \bibnamefont
  {Peeters}},\ and\ \bibinfo {author} {\bibfnamefont {A.~V.}\ \bibnamefont
  {Vagov}},\ }\bibfield  {title} {\bibinfo {title} {Extended ginzburg-landau
  formalism for two-band superconductors},\ }\href
  {https://doi.org/10.1103/PhysRevLett.106.047005} {\bibfield  {journal}
  {\bibinfo  {journal} {Phys. Rev. Lett.}\ }\textbf {\bibinfo {volume} {106}},\
  \bibinfo {pages} {047005} (\bibinfo {year} {2011})}\BibitemShut {NoStop}%
\bibitem [{\citenamefont {Vagov}\ \emph {et~al.}(2016)\citenamefont {Vagov},
  \citenamefont {Shanenko}, \citenamefont {Milo{\v{s}}evi{\'c}}, \citenamefont
  {Axt}, \citenamefont {Vinokur}, \citenamefont {Aguiar},\ and\ \citenamefont
  {Peeters}}]{vagov16}%
  \BibitemOpen
  \bibfield  {author} {\bibinfo {author} {\bibfnamefont {A.}~\bibnamefont
  {Vagov}}, \bibinfo {author} {\bibfnamefont {A.~A.}\ \bibnamefont {Shanenko}},
  \bibinfo {author} {\bibfnamefont {M.}~\bibnamefont {Milo{\v{s}}evi{\'c}}},
  \bibinfo {author} {\bibfnamefont {V.~M.}\ \bibnamefont {Axt}}, \bibinfo
  {author} {\bibfnamefont {V.~M.}\ \bibnamefont {Vinokur}}, \bibinfo {author}
  {\bibfnamefont {J.~A.}\ \bibnamefont {Aguiar}},\ and\ \bibinfo {author}
  {\bibfnamefont {F.~M.}\ \bibnamefont {Peeters}},\ }\bibfield  {title}
  {\bibinfo {title} {Superconductivity between standard types: Multiband versus
  single-band materials},\ }\href {https://doi.org/10.1103/PhysRevB.93.174503}
  {\bibfield  {journal} {\bibinfo  {journal} {Phys. Rev. B}\ }\textbf {\bibinfo
  {volume} {93}},\ \bibinfo {pages} {174503} (\bibinfo {year}
  {2016})}\BibitemShut {NoStop}%
\bibitem [{\citenamefont {de~Oliveira}(2017)}]{isaias17}%
  \BibitemOpen
  \bibfield  {author} {\bibinfo {author} {\bibfnamefont {I.~G.}\ \bibnamefont
  {de~Oliveira}},\ }\bibfield  {title} {\bibinfo {title} {The threshold
  temperature where type-{I} and type-{II} interchange in mesoscopic
  superconductors at the bogomolnyi limit},\ }\href
  {https://doi.org/https://doi.org/10.1016/j.physleta.2017.01.032} {\bibfield
  {journal} {\bibinfo  {journal} {Phys. Lett. A}\ }\textbf {\bibinfo {volume}
  {381}},\ \bibinfo {pages} {1248 } (\bibinfo {year} {2017})}\BibitemShut
  {NoStop}%
\bibitem [{\citenamefont {C\'ordoba-Camacho}\ \emph {et~al.}(2016)\citenamefont
  {C\'ordoba-Camacho}, \citenamefont {da~Silva}, \citenamefont {Vagov},
  \citenamefont {Shanenko},\ and\ \citenamefont {Aguiar}}]{camacho2016}%
  \BibitemOpen
  \bibfield  {author} {\bibinfo {author} {\bibfnamefont {W.~Y.}\ \bibnamefont
  {C\'ordoba-Camacho}}, \bibinfo {author} {\bibfnamefont {R.~M.}\ \bibnamefont
  {da~Silva}}, \bibinfo {author} {\bibfnamefont {A.}~\bibnamefont {Vagov}},
  \bibinfo {author} {\bibfnamefont {A.~A.}\ \bibnamefont {Shanenko}},\ and\
  \bibinfo {author} {\bibfnamefont {J.~A.}\ \bibnamefont {Aguiar}},\ }\bibfield
   {title} {\bibinfo {title} {Between types {I} and {II}: Intertype flux exotic
  states in thin superconductors},\ }\href
  {https://doi.org/10.1103/PhysRevB.94.054511} {\bibfield  {journal} {\bibinfo
  {journal} {Phys. Rev. B}\ }\textbf {\bibinfo {volume} {94}},\ \bibinfo
  {pages} {054511} (\bibinfo {year} {2016})}\BibitemShut {NoStop}%
\bibitem [{\citenamefont {C\'ordoba-Camacho}\ \emph {et~al.}(2018)\citenamefont
  {C\'ordoba-Camacho}, \citenamefont {da~Silva}, \citenamefont {Vagov},
  \citenamefont {Shanenko},\ and\ \citenamefont {Aguiar}}]{camacho2018}%
  \BibitemOpen
  \bibfield  {author} {\bibinfo {author} {\bibfnamefont {W.~Y.}\ \bibnamefont
  {C\'ordoba-Camacho}}, \bibinfo {author} {\bibfnamefont {R.~M.}\ \bibnamefont
  {da~Silva}}, \bibinfo {author} {\bibfnamefont {A.}~\bibnamefont {Vagov}},
  \bibinfo {author} {\bibfnamefont {A.~A.}\ \bibnamefont {Shanenko}},\ and\
  \bibinfo {author} {\bibfnamefont {J.~A.}\ \bibnamefont {Aguiar}},\ }\bibfield
   {title} {\bibinfo {title} {Quasi-one-dimensional vortex matter in
  superconducting nanowires},\ }\href
  {https://doi.org/10.1103/PhysRevB.98.174511} {\bibfield  {journal} {\bibinfo
  {journal} {Phys. Rev. B}\ }\textbf {\bibinfo {volume} {98}},\ \bibinfo
  {pages} {174511} (\bibinfo {year} {2018})}\BibitemShut {NoStop}%
\bibitem [{\citenamefont {Cadorim}\ \emph {et~al.}(2019)\citenamefont
  {Cadorim}, \citenamefont {de~Oliveira~Calsolari}, \citenamefont {Zadorosny},\
  and\ \citenamefont {Sardella}}]{cadorim2019}%
  \BibitemOpen
  \bibfield  {author} {\bibinfo {author} {\bibfnamefont {L.~R.}\ \bibnamefont
  {Cadorim}}, \bibinfo {author} {\bibfnamefont {T.}~\bibnamefont
  {de~Oliveira~Calsolari}}, \bibinfo {author} {\bibfnamefont {R.}~\bibnamefont
  {Zadorosny}},\ and\ \bibinfo {author} {\bibfnamefont {E.}~\bibnamefont
  {Sardella}},\ }\bibfield  {title} {\bibinfo {title} {Crossover from type {I}
  to type {II} regime of mesoscopic superconductors of the first group},\
  }\href {https://doi.org/10.1088/1361-648x/ab4a4a} {\bibfield  {journal}
  {\bibinfo  {journal} {J. Phys. Condens. Matter}\ }\textbf {\bibinfo {volume}
  {32}},\ \bibinfo {pages} {095304} (\bibinfo {year} {2019})}\BibitemShut
  {NoStop}%
\bibitem [{\citenamefont {C{\'{o}}rdoba-Camacho}\ \emph
  {et~al.}(2019)\citenamefont {C{\'{o}}rdoba-Camacho}, \citenamefont
  {da~Silva}, \citenamefont {Shanenko}, \citenamefont {Vagov}, \citenamefont
  {Vasenko}, \citenamefont {Lvov},\ and\ \citenamefont {Aguiar}}]{camacho2020}%
  \BibitemOpen
  \bibfield  {author} {\bibinfo {author} {\bibfnamefont {W.~Y.}\ \bibnamefont
  {C{\'{o}}rdoba-Camacho}}, \bibinfo {author} {\bibfnamefont {R.~M.}\
  \bibnamefont {da~Silva}}, \bibinfo {author} {\bibfnamefont {A.~A.}\
  \bibnamefont {Shanenko}}, \bibinfo {author} {\bibfnamefont {A.}~\bibnamefont
  {Vagov}}, \bibinfo {author} {\bibfnamefont {A.~S.}\ \bibnamefont {Vasenko}},
  \bibinfo {author} {\bibfnamefont {B.~G.}\ \bibnamefont {Lvov}},\ and\
  \bibinfo {author} {\bibfnamefont {J.~A.}\ \bibnamefont {Aguiar}},\ }\bibfield
   {title} {\bibinfo {title} {Spontaneous pattern formation in superconducting
  films},\ }\href {https://doi.org/10.1088/1361-648x/ab5379} {\bibfield
  {journal} {\bibinfo  {journal} {J. Phys. Condens. Matter}\ }\textbf {\bibinfo
  {volume} {32}},\ \bibinfo {pages} {075403} (\bibinfo {year}
  {2019})}\BibitemShut {NoStop}%
\bibitem [{\citenamefont {Gropp}\ \emph {et~al.}(1996)\citenamefont {Gropp},
  \citenamefont {Kaper}, \citenamefont {Leaf}, \citenamefont {Levine},
  \citenamefont {Palumbo},\ and\ \citenamefont {Vinokur}}]{gropp1996}%
  \BibitemOpen
  \bibfield  {author} {\bibinfo {author} {\bibfnamefont {W.~D.}\ \bibnamefont
  {Gropp}}, \bibinfo {author} {\bibfnamefont {H.~G.}\ \bibnamefont {Kaper}},
  \bibinfo {author} {\bibfnamefont {G.~K.}\ \bibnamefont {Leaf}}, \bibinfo
  {author} {\bibfnamefont {D.~M.}\ \bibnamefont {Levine}}, \bibinfo {author}
  {\bibfnamefont {M.}~\bibnamefont {Palumbo}},\ and\ \bibinfo {author}
  {\bibfnamefont {V.~M.}\ \bibnamefont {Vinokur}},\ }\bibfield  {title}
  {\bibinfo {title} {Numerical simulation of vortex dynamics in type-{II}
  superconductors},\ }\href
  {https://doi.org/https://doi.org/10.1006/jcph.1996.0022} {\bibfield
  {journal} {\bibinfo  {journal} {J. Comp. Phys.}\ }\textbf {\bibinfo {volume}
  {123}},\ \bibinfo {pages} {254 } (\bibinfo {year} {1996})}\BibitemShut
  {NoStop}%
\bibitem [{\citenamefont {Zharkov}\ \emph {et~al.}(2000)\citenamefont
  {Zharkov}, \citenamefont {Zharkov},\ and\ \citenamefont
  {Zvetkov}}]{zharkov00}%
  \BibitemOpen
  \bibfield  {author} {\bibinfo {author} {\bibfnamefont {G.~F.}\ \bibnamefont
  {Zharkov}}, \bibinfo {author} {\bibfnamefont {V.~G.}\ \bibnamefont
  {Zharkov}},\ and\ \bibinfo {author} {\bibfnamefont {A.~Y.}\ \bibnamefont
  {Zvetkov}},\ }\bibfield  {title} {\bibinfo {title} {Ginzburg-landau
  calculations for a superconducting cylinder in a magnetic field},\ }\href
  {https://doi.org/10.1103/PhysRevB.61.12293} {\bibfield  {journal} {\bibinfo
  {journal} {Phys. Rev. B}\ }\textbf {\bibinfo {volume} {61}},\ \bibinfo
  {pages} {12293} (\bibinfo {year} {2000})}\BibitemShut {NoStop}%
\bibitem [{\citenamefont {Zharkov}(2001)}]{zharkov01}%
  \BibitemOpen
  \bibfield  {author} {\bibinfo {author} {\bibfnamefont {G.~F.}\ \bibnamefont
  {Zharkov}},\ }\bibfield  {title} {\bibinfo {title} {Paramagnetic meissner
  effect in superconductors from self-consistent solution of ginzburg-landau
  equations},\ }\href {https://doi.org/10.1103/PhysRevB.63.214502} {\bibfield
  {journal} {\bibinfo  {journal} {Phys. Rev. B}\ }\textbf {\bibinfo {volume}
  {63}},\ \bibinfo {pages} {214502} (\bibinfo {year} {2001})}\BibitemShut
  {NoStop}%
\bibitem [{\citenamefont {Zharkov}(2002)}]{zharkov02}%
  \BibitemOpen
  \bibfield  {author} {\bibinfo {author} {\bibfnamefont {G.~F.}\ \bibnamefont
  {Zharkov}},\ }\bibfield  {title} {\bibinfo {title} {On the emergence of
  superconductivity and hysteresis in a cylindrical type {I} superconductor},\
  }\href {https://doi.org/10.1134/1.1513826} {\bibfield  {journal} {\bibinfo
  {journal} {Journal of Experimental and Theoretical Physics}\ }\textbf
  {\bibinfo {volume} {95}},\ \bibinfo {pages} {517} (\bibinfo {year}
  {2002})}\BibitemShut {NoStop}%
\bibitem [{\citenamefont {Zharkov}(2003)}]{zharkov03}%
  \BibitemOpen
  \bibfield  {author} {\bibinfo {author} {\bibfnamefont {G.~F.}\ \bibnamefont
  {Zharkov}},\ }\bibfield  {title} {\bibinfo {title} {First and second order
  phase transitions and magnetic hysteresis in a superconducting plate},\
  }\href {https://doi.org/10.1023/A:1021845418088} {\bibfield  {journal}
  {\bibinfo  {journal} {Journal of Low Temperature Physics}\ }\textbf {\bibinfo
  {volume} {130}},\ \bibinfo {pages} {45} (\bibinfo {year} {2003})}\BibitemShut
  {NoStop}%
\end{thebibliography}%

\end{document}